\documentclass[12pt,letterpaper,copyedit]{article}
\usepackage{ol2}
\usepackage[draft]{hyperref}
\begin{document}
\title{Polarization-Current-Based FDTD Near-to-Far-Field Transformation}
\author{Yong Zeng and Jerome V. Moloney}
\address{Arizona Center for mathematical Sciences, University of
Arizona, Tucson, Arizona 85721\\
$^*$Corresponding author: zengy@acms.arizona.edu}
\begin{abstract}
A new near-to-far-field transformation algorithm for
three-dimensional finite-different time-domain is presented in
this article. This new approach is based directly on the
polarization current of the scatterer, not the scattered near
fields. It therefore eliminates the numerical errors originating
from the spatial offset of the E and H fields, inherent in the
standard near-to-far-field transformation. The proposed method is
validated via direct comparisons with the analytical Lorentz-Mie
solutions of plane waves scattered by large dielectric and
metallic spheres with strong forward-scattering lobes.
\end{abstract}

The grid-based finite-difference time-domain (FDTD) method is one
of the most popular Maxwell solvers, which has been proven to be
efficient, stable and easy-to-implement \cite{taflove}. Due to the
computational resource limitation, a FDTD simulation truncates the
open boundary to a spatial domain adjacent to the scatterer. The
near-to-far-field (NTFF) transformation, therefore, is routinely
employed to obtain the far-zone information such as antenna
scattering patterns and nanocavity radiation patterns
\cite{taflove,drezek,guo,brock,oetting,kim,onuta,husnik}. The
standard NTFF (S-NTFF) transformation, introduced in the early
1980s, is based on the vector Kirchhoff integral relation
\cite{taflove}. The scattered fields in the far zone are
calculated through an integration of the near-zone fields over a
virtual closed (Huygens) surface completely enclosing the
scatterer \cite{jackson}. To accomplish this we need compute, via
FDTD and discrete Fourier transformation, the scattered E and H
fields tangential to the fictitious surface. However, the spatial
and temporal offset between E and H field, a character of the Yee
update scheme \cite{taflove}, may result in unacceptable numerical
errors \cite{sun,zhai,yurkin}. For instance, Ref.\cite{li}
demonstrated that the accuracy of the S-NTFF is unacceptable when
calculating the backscattering from strongly forward-scattering
objects, and the relative error may be as high as two orders of
magnitude at short wavelength. To improve its numerical accuracy,
at least two different modifications have been proposed, including
discarding information in the forward-scattering region \cite{li}
and using geometric mean in place of arithmetic mean \cite{dirk}.

In classical electrodynamics, the optical response of medium made
up of a large number of atoms or molecules originates from the
perturbed motions of the charges bound in each molecule. The
molecule charge density is distorted by the external
electromagnetic fields, and further produces an electric
polarization $\mathbf{P}$ in the medium \cite{jackson}.
Consequently in this article we proposal using the polarization
current, rather than the scattered near fields, to derive the
far-zone information. The proposed approach is validated via
comparison with rigorous analytical solutions of plane wave
scattered by large dielectric and metallic spheres. A detailed
discussion regarding its advantages and disadvantages is also
presented.

In the absence of sources, the Maxwell equations are read as
\begin{eqnarray}
&&\nabla\cdot\mathbf{B}=0,\:\:\:\:\:\:
\nabla\times\mathbf{E}=-\frac{\partial\mathbf{B}}{\partial
t},\:\cr
&&\nabla\cdot\mathbf{D}=0,\:\:\:\:\:\:\nabla\times\mathbf{H}=\frac{\partial\mathbf{D}}{\partial
t}.
\end{eqnarray}
Assuming the following constitutive relations
$\mathbf{D}=\epsilon_{0}\mathbf{E}+\mathbf{P}$ as well as
$\mathbf{B}=\mu_{0}\mathbf{H}$ (the scatterer is therefore
nonmagnetic), we arrive an inhomogeneous Helmholtz wave equation
for the vector potential $\mathbf{A}$ in the Lorentz gauge
\cite{jackson}
\begin{equation}
\nabla^{2}\mathbf{A}-\frac{1}{c^{2}}\frac{\partial^{2}
\mathbf{A}}{\partial t^{2}}=-\mu_{0}\mathbf{J}(t), \label{eq1}
\end{equation}
where $\mathbf{J}$ is the polarization current defined as
$\mathbf{J}=\partial\mathbf{P}/\partial t$ \cite{neil}. With a
time dependence $e^{-i\omega t}$ understood, Eq.~(\ref{eq1})
becomes
\begin{equation}
\nabla^{2}\mathbf{A}+k^{2}\mathbf{A}=-\mu_{0}\mathbf{J}(\omega),
\end{equation}
with $k=\omega/c$ being the vacuum wave number. A formal solution
of the above equation, with the help of free-space Green function,
can be written as \cite{jackson}
\begin{equation}
\mathbf{A}(\mathbf{r})-\mathbf{A}^{(0)}(\mathbf{r})=\frac{\mu_{0}}{4\pi}\int_{v}\mathbf{J}(\mathbf{r'})\frac{e^{ik|\mathbf{r}-\mathbf{r'}|}}{|\mathbf{r}-\mathbf{r'}|}d\mathbf{r'}.
\end{equation}
The left-hand side represents the scattered wave with
$\mathbf{A}^{(0)}$ representing the incident wave. To obtain the
field in the radiation zone, it is sufficient to approximate the
numerator with \cite{jackson}
\begin{equation}
|\mathbf{r}-\mathbf{r'}|\approx r-\mathbf{n}\cdot\mathbf{r'},
\end{equation}
while in the denominator $|\mathbf{r}-\mathbf{r'}|\approx r$. Here
$\mathbf{n}$ is a unit vector in the direction of $\mathbf{r}$.
The far-zone vector potential of the scattered field is therefore
expressed as
\begin{equation}
\lim_{r\rightarrow\infty}\mathbf{A}(\mathbf{r})\approx\frac{\mu_{0}e^{ikr}}{4\pi
r}\int_{v}\mathbf{J}(\mathbf{r'})e^{-ik\mathbf{n}\cdot\mathbf{r'}}d\mathbf{r'}\equiv
\frac{\mu_{0}e^{ikr}}{4\pi r}\mathbf{p}.
\end{equation}
It behaves as an outgoing spherical wave while depends on the
polar and azimuth angles ($\theta$,$\phi$) in spherical polar
coordinate. The scattered fields in the radiation zone, keeping
only the leading order, are further given as
\begin{equation}
\mathbf{B}=\frac{ik\mu_{0}}{4\pi
r}e^{ikr}\mathbf{n}\times\mathbf{p},\:\:\:\:\:\: \mathbf{E}=
c\mathbf{B}\times\mathbf{n}. \label{eq3}
\end{equation}
Evidently, both fields are transverse to the radius vector
$\mathbf{n}$ and fall off as $r^{-1}$.

The time-averaged power scattered per unit solid angle is
\begin{equation}
\frac{dP}{d\Omega}=\frac{1}{2}\textrm{Re}\left[r^{2}\mathbf{n}\cdot\mathbf{E}\times\mathbf{H}^{\ast}\right]=\frac{k^{2}\eta_{0}
}{32\pi^2}\left|(\mathbf{n}\times\mathbf{p})\times\mathbf{n}\right|^{2}
\label{eq4}
\end{equation}
with $\eta_{0}=\sqrt{\mu_{0}/\epsilon_{0}}$ being the intrinsic
impedance of free space. To obtain the scattering cross section,
the scattered power needs be normalized to the incident power
\cite{jackson}. It is worth noting that the above general
expression is quite similar to that of an oscillating electric or
magnetic dipole, except that the $\mathbf{p}(\mathbf{n})$ here is
angular-dependent \cite{jackson}.

\begin{figure}[t]
\centering
\includegraphics[width=0.5\textwidth]{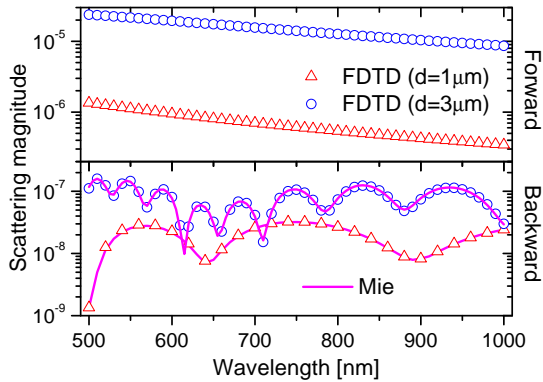}\vspace*{-6.5cm}
\caption{Forward and backward scattering magnitudes of 1-$\mu$m
and 3-$\mu$m diameter dielectric spheres with $\epsilon_{r}=1.21$.
Both numerical FDTD solution and rigorous Lorentz-Mie solution are
presented. The cell size employed in FDTD simulation is 25 $n$m.
The corresponding  S-NTFF results can be found from Fig.1 of
Ref.\cite{li}.} \label{fig1}
\end{figure}

The above derivation immediately implies two features of our
approach: (1) It depends only on the polarization current
$\mathbf{J}$ but not the scattered electromagnetic fields
$\mathbf{E}$ and $\mathbf{H}$, the error induced by the spatial
offset between them in the S-NTFF is therefore completely avoided;
(2) Unlike the S-NTFF, there is no requirement of scattered
fields. Consequently, our approach is easier to implement for
strongly convergent light sources such as Gaussian beam
\cite{jean}.

It should be emphasized that we can very easily extract the
polarization current from FDTD, because an auxiliary differential
equation of the current is frequently adopted by standard FDTD to
simulate dispersive, nonlinear and even quantum-mechanical media
\cite{taflove}. For the special case of a dielectric scatterer,
the standard FDTD itself is enough to provide the information of
the current, since it connects to the electric field as $
\mathbf{J}(\omega)=-i\omega\epsilon_{0}\chi\mathbf{E}(\omega)$
with $\chi$ being the susceptibility.

To validate our approach as well as the corresponding numerical
algorithm, we begin by considering the backscattering from two
strongly forward-scattering dielectric spheres with a relative
permittivity $\epsilon_{r}=1.21$ ($\chi=0.21$) which were
considered in Ref.\cite{li,dirk} with the use of the (modified)
S-NTFF. Assuming the incident field propagates along the positive
$x$ direction, the backscattering (corresponding to
$\mathbf{n}=(-1,0,0)$) power is simply given as
\begin{equation}
\frac{k^{2}\eta_{0}}{32\pi^2}\left[|p_{y}(\mathbf{n})|^{2}+|p_{z}(\mathbf{n})|^{2}\right],
\label{eq5}
\end{equation}
that is, we do not need consider the $x$-component current. In our
simulations, the FDTD lattice is terminated by a uniaxial
perfectly matched layer (UMPL) in all directions \cite{taflove},
and the whole computational space is divided into uniform cubic
cells with size of 25 $n$m. Furthermore, an impulsive wideband
plane wave is excited to obtain the spectra in the wavelength
domain 500-1000 $n$m.

Fig.1 presents the FDTD calculated far-field spectra for two
spheres with diameters $d=1 \mu$m and $d=3 \mu$m. Because their
sizes are on the order of incident wavelength, the appearance of
higher-order multipoles distort the symmetrical electric-dipole
pattern and result in strong forward scattering. The biggest
contrast between forward and backward scattering appears for the
1$\mu$m-diameter sphere. At 500-$n$m wavelength, its scattering in
the forward direction is almost three orders of magnitude stronger
than its backward counterpart. As a reference, the analytical
Lorentz-Mie solutions are also shown. Clearly, excellent agreement
is achieved even when the incident wavelength is coarsely resolved
on the FDTD grid. Under the same conditions, the relative error of
the S-NTFF result is as high as two orders of magnitude \cite{li},
and modifications such as the geometric mean must be introduced to
improve its numerical accuracy \cite{dirk}.

We further consider the scattering of a 1$\mu$m-diameter gold
sphere with Drude-type permittivity approximated as
\begin{equation}
\epsilon(\omega)=1.0-\frac{\omega_{p}^{2}}{\omega(\omega+i\gamma)}
\end{equation}
in the wavelength region 700-1200 $n$m. The bulk plasma frequency
$\omega_{p}$ is taken as $1.367\times 10^{16}$$s^{-1}$, and the
phenomenological collision frequency $\gamma=6.478\times
10^{13}$$s^{-1}$ \cite{palik}. Because of the negative
permittivity, surface plasmon polaritons (SPP) are excited and the
resulting evanescent waves concentrate around the metallic
surface. To simulate these nonpropagating fields, sufficiently
finer resolution is needed in FDTD to achieve acceptable accuracy.
Fig.2 shows the comparison of the analytical Lorentz-Mie solution
with our FDTD numerical results (with cell size of 10 $n$m).
Clearly, they are quite close, except tiny blue shifts of SPP
resonances as well as a relative difference of roughly 5\% around
the 770-$n$m wavelength. It should be emphasized that these
disagreements are induced by the inherent staircasing of the
standard FDTD on a Cartesian grid and not due to our NTFF. In the
simulation the diameter of the metallic sphere is resolved by 100
cells, and better agreement can be achieved by decreasing the cell
size.

The key differences between our current-based NTFF with the S-NTFF
can be summarized as follow: (1) Origins of numerical errors are
different. They both suffer from staircasing. However, the S-NTFF
has two more sources including the scattered-field requirement (we
thus need total field/scattered field or pure scattered field
scheme \cite{taflove}) and the spatial offset of the E and H
fields over the integral surface \cite{li,dirk}; (2) A virtual
integral surface enclosing the whole scatterers is not needed in
our approach. It is therefore easily and efficiently implemented
within the framework of standard FDTD codes; (3) The two methods
have different computational burdens. Consider a cubic scatterer
resolved by $\rm{N}^{3}$ cells, the total number of discrete
Fourier transformations to be performed is $3\times\rm{N}^{3}$ in
volume integration (our method) in comparison with
$12\times\rm{N}^{2}$ in surface integration (S-NTFF). On the
contrary, a multiple scatterer cluster with small volume but large
surface may require more calculational resources in the S-NTFF;
(4) The current-based NTFF has a close relation with the
analytical multipole expansion, it is therefore convenient to
identify the contributions from different multipolar sources and
further provide insights into complex electromagnetic processes
including radiation, diffraction and scattering
\cite{oetting,jackson}.

\begin{figure}[t]
\centering
\includegraphics[width=0.5\textwidth]{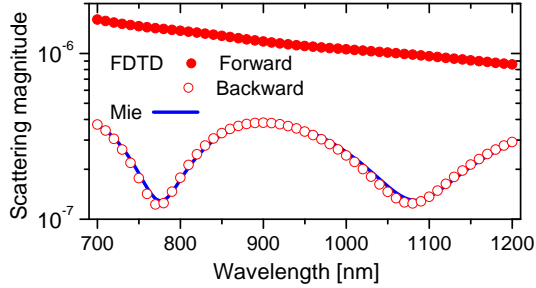}\vspace*{-7.8cm}
\caption{Far-field scattering spectra of 1$\mu$m-diameter gold
sphere. Both numerical FDTD solution and rigorous Lorentz-Mie
solution are presented. The cell size employed in FDTD simulation
is 10 $n$m.} \label{fig2}
\end{figure}

It should be pointed out that the spatial offset between E and H
field can be avoided by using the scalar Kirchhoff integral
formula
\begin{equation}
\psi(\mathbf{r})=-\frac{1}{4\pi}\int_{\rm{S}}\frac{e^{ik\rm{R}}}{\rm{R}}\left[\nabla'\psi+ik\left(1+\frac{i}{k\rm{R}}\right)\frac{\mathbf{R}}{\rm{R}}\psi\right]\cdot
d\mathbf{s}
\end{equation}
with $\mathbf{R}=\mathbf{r}-\mathbf{r'}$ and $\psi$ being any
component of $\mathbf{E}$ or $\mathbf{H}$ \cite{jackson}. However,
the integral surface $\rm{S}$ generally is infinite for an
open-boundary problem. To achieve acceptable accuracy in numerical
simulation, we need to truncate it to a finite surface with
sufficient area. Consider an electric dipole with size $a$. The
radius of the integral surface should be roughly $10a$ to contain
the field with amplitude bigger than 10\% of the dipole magnitude.
It thus increases the computational burden.

To conclude, a polarization-current-based near-to-far-field
transformation for the three-dimensional finite-difference
time-domain algorithm is developed. Because it completely
eliminates the spatial-offset error inherent in the surface
integration of standard near-to-far-field scheme, the new method
can achieve improved accuracy even with coarse grids. Its validity
and efficiency is further demonstrated via direct comparisons with
analytical theory by calculating the backscattering from strongly
forward-scattering objects including dielectric and metallic
spheres. We also compare it to the standard near-to-far-field
transformation in terms of calculational cost and accuracy. The
new method may have potential applications in biophotonics and
nanophotonics.

The authors thank Prof. Masud Mansuripur, Prof. Moysey Brio and
Dr. Colm Dineen for their invaluable discussions. This work is
supported by the Air Force Office of Scientific Research (AFOSR),
under Grant No. FA9550-07-1-0010 and FA9550-04-1-0213. J. V.
Moloney acknowledges support from the Alexander von Humboldt
foundation.


\begin{thebibliography}{99}
\bibitem{taflove} A. Taflove and S. C. Hagness, {\it Computational Electrodynamics: the finite-difference time-domain method} (Third Edition, Artech House, Boston, 2005).
\bibitem{drezek} R. Drezek, A. Dunn, and R. Richards-Kortum, ``Light scattering from cells: Finite-difference time-domain simulations and goniometric measurements," Appl. Opt. 38, 3651 (1999).
\bibitem{guo} S. Guo and S. Albin, ``Numerical techniques for excitation and analysis of defect modes in photonic crystals," Opt. Express 11, 1080
(2003).
\bibitem{brock} R. Scott Brock, X-H Hu, P. Yang, J. Lu, ``Evaluation of a parallel FDTD code and application to modeling of light scattering by deformed red blood
cells," Opt. Express 13, 5279 (2005).
\bibitem{oetting} C-C Oetting, L. Klinkenbusch, ``Near-to-far-field transformation by a time-domain spherical-multipole
analysis," IEEE Trans. Antennas Propag. 53, 2054 (2005).
\bibitem{kim} S-H Kim, S-K Kim, Y-H Lee, ``Vertical beaming of
wavelength-scale photonic crystal resonators," Phys. Rev. B 73,
235117 (2006).
\bibitem{onuta} T-D Onuta, M. Waegele, C. C. DuFort, W. L. Schaich, B.
Dragnea, ``Optical Field Enhancement at Cusps between Adjacent
Nanoapertures," Nano Lett. 7, 557 (2007).
\bibitem{husnik} M. Husnik, M. W. Klein, N. Feth, M. K\"{o}nig, J. Niegemann, K. Busch,
S. Linden, and M. Wegener, ``Absolute Extinction Cross Section of
Individual Magnetic Split-Ring Resonators," Nature Photon. 2, 614
(2008).
\bibitem{jackson} J. D. Jackson, {\it Classical Electrodynamics},
Second Edition, John Wiley \& Sons, 1975.
\bibitem{sun} W. Sun, Q. Fu, ``Finite-Difference Time-Domain Solution of Light Scattering by Dielectric Particles with Large Complex Refractive
Indices," Appl. Opt. 39, 5569 (2000).
\bibitem{zhai} P. W. Zhai, Y. K. Lee, G. W. Kattawar, and P. Yang, ``Implementing the near to far-field transformation in the finite-difference time-domain method," Appl. Opt. 43, 3738, (2004).
\bibitem{yurkin} M. A. Yurkin, A. G. Hoekstra, R. S. Brock, Jun Q. Lu,
``Systematic comparison of the discrete dipole approximation and
the finite difference time domain method for large dielectric
scatterers," Opt. Express 15, 17902 (2007).
\bibitem{li} X. Li, A. Taflove, V. Backman, ``Modified FDTD near-to-far-field transformation for improved backscattering calculation
of strongly forward-scattering objects," IEEE Antennas Wireless
Propag. Lett. 4, 35 (2005).
\bibitem{dirk} D. J. Robinson, J. B. Schneider ``On the use of the geometric mean in FDTD near-to-far-field transformations,"IEEE Antennas Wireless
Propag. Lett. 55, 3204 (2007).
\bibitem{neil} N. V. Budko, ``Observation of Locally Negative Velocity of the Electromagnetic Field in Free
Space," Phys. Rev. Lett. 102, 020401 (2009).
\bibitem{jean} J. Lerm\'{e}, C. Bonnet, M. Broyer, E. Cottancin, S. Marhaba, M.
Pellarin, ``Optical response of metal or dielectric nano-objects
in strongly convergent light beams," Phys. Rev. B 77, 245406
(2008).
\bibitem{palik} E. D. Palik, {\it Handbook of optical constants of solids},
Academic, Orlando, 1985.
\end{thebibliography}

\newpage
\end{document}